# Requirements for Power Converters

*J.-P. Burnet*
CERN, Geneva, Switzerland

**Abstract**
This paper introduces the requirements for power converters needed for particle accelerators. It describes the role of power converters and the challenges and constraints when powering magnets. The different circuit layouts are presented as well as the operating cycles. The power converter control and high precision definition are also introduced. This paper lists the key circuit parameters to be taken into consideration to properly specify a power converter that can be compiled in a functional specification.

## 1  Introduction

Particle accelerators are very special machines that consume a lot of electricity. The main devices consuming this energy are the magnets and the radio-frequency (RF) systems. It represents from 70% to 90% of the total energy needed for a machine. The way to power these two main loads is always special and needs a lot of care to obtain the right performance from the machine, see Fig. 1. The performance of the power converters directly impacts upon the beam quality, and the requirements for power converters are outside the range of classical industrial products. This paper will introduce the power converters needed for particle accelerators, and their main parameters.

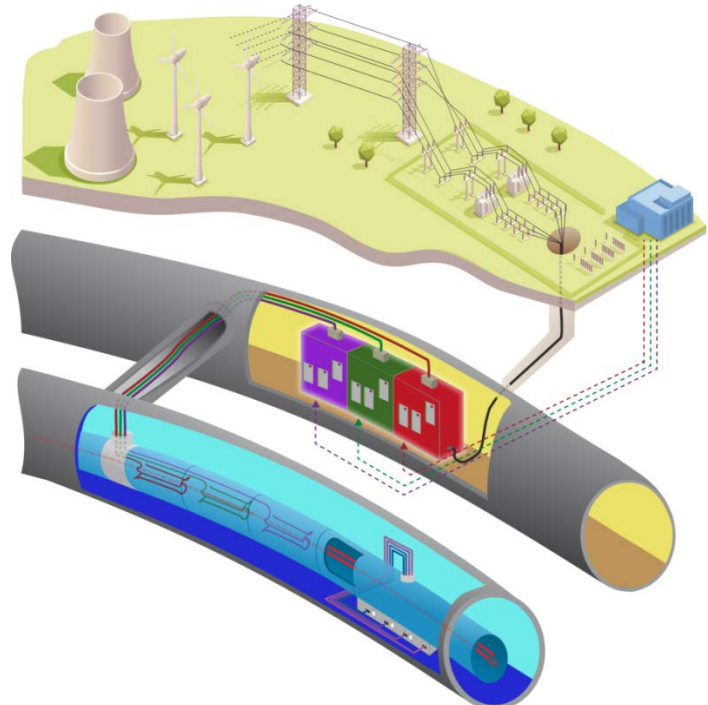

**Fig. 1:** Particle accelerator powering chain

## 2    Type of loads

The most popular particle accelerator today is the synchrotron machine. In such machines, a large number of magnets is used to control the beam and, depending of the type of particle, there may be many radio-frequency cavities for acceleration, see Fig. 2. This paper will mainly concentrate upon magnet powering, as radio-frequency powering is covered by another paper.

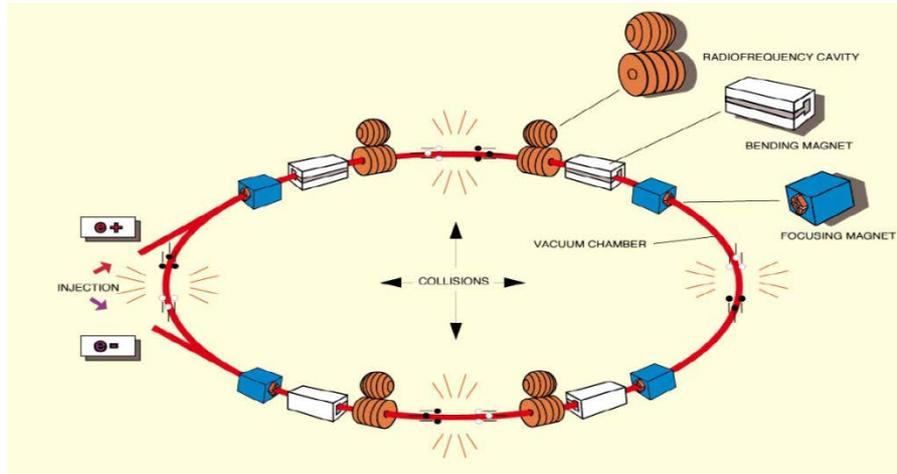

**Fig. 2:** Basic concept of a synchrotron machine

The power converters or power supplies are the devices that take energy from the electrical grid and transform it to control the flow of energy supplied to the magnets or radio-frequency power amplifiers. The magnet power converters are always of an AC/DC type.

## 3    Powering magnets

Different types of magnets are needed for particle accelerators, see Fig. 3. The two main families are the dipole magnets that bend the beam and the quadrupole magnets that focus the beam. Sextupole magnets are used to correct the chromaticity, and octupole magnets control the Landau damping. Particle accelerators also need a lot of small correctors to compensate for local distortion due to magnet imperfections, alignment, etc.

A synchrotron is composed of a periodic repetition of focus-defocus (FODO) cells. Each cell contains dipole, quadrupole and sextupole magnets, see Figs. 4 and 5 [1].

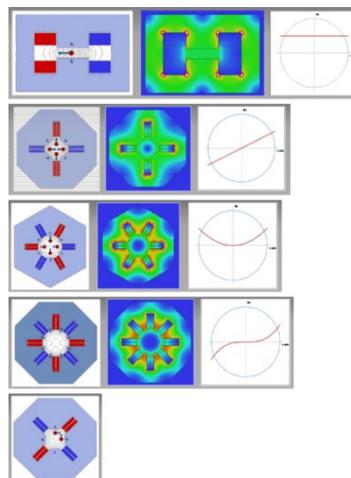

**Fig. 3:** Different magnet types: (a) dipole magnet; (b) quadrupole magnet; (c) sextupole magnet; (d) octupole magnet.

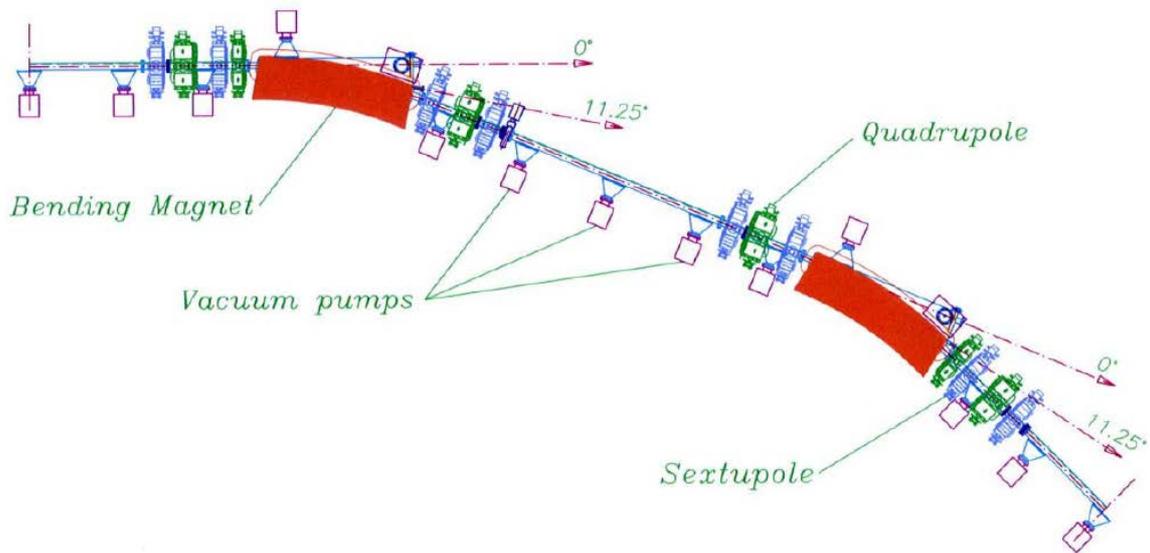

**Fig. 4:** Arrangement of the magnets within one unit cell of the SESAME storage ring

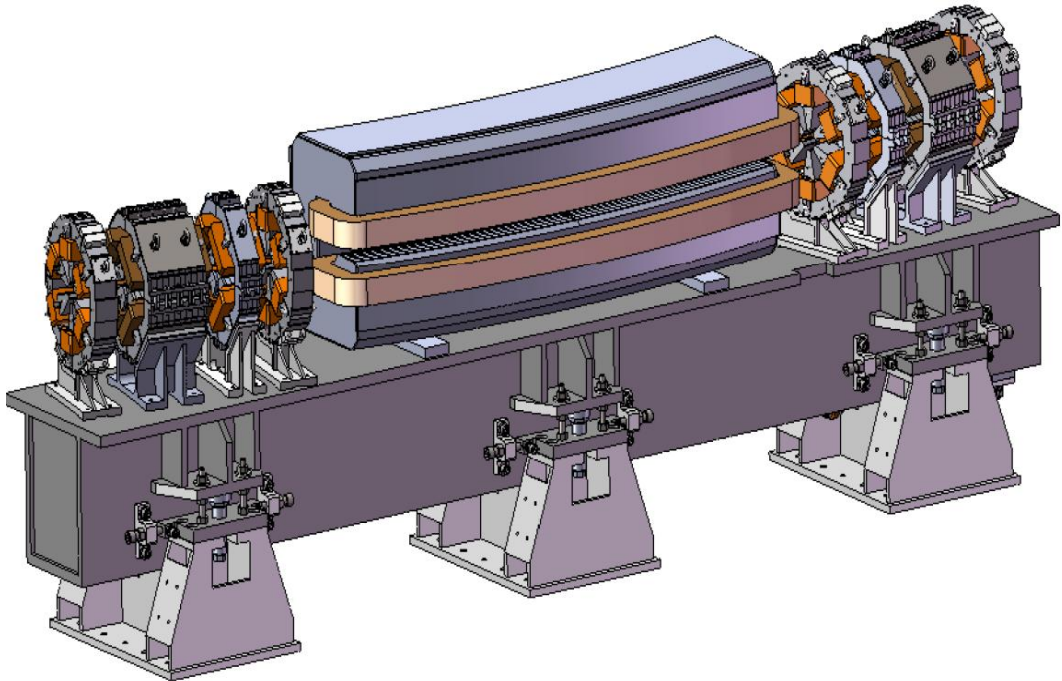

**Fig. 5:** SESAME FODO cell

The magnets are made with a magnetic core (laminated or not) and winding coils. As seen by the power converters, the magnets are always inductive loads. The equivalent circuit is an inductance in series with a resistor (due to the resistance of the coils and the DC cables between the power converter and the magnet).

In a synchrotron the beam energy is proportional to the magnetic field of the dipole magnets. The magnetic field is generated by the current circulating in the magnet coils, see Fig. 6. To then control the beam, the operators need to control the current through the magnets.

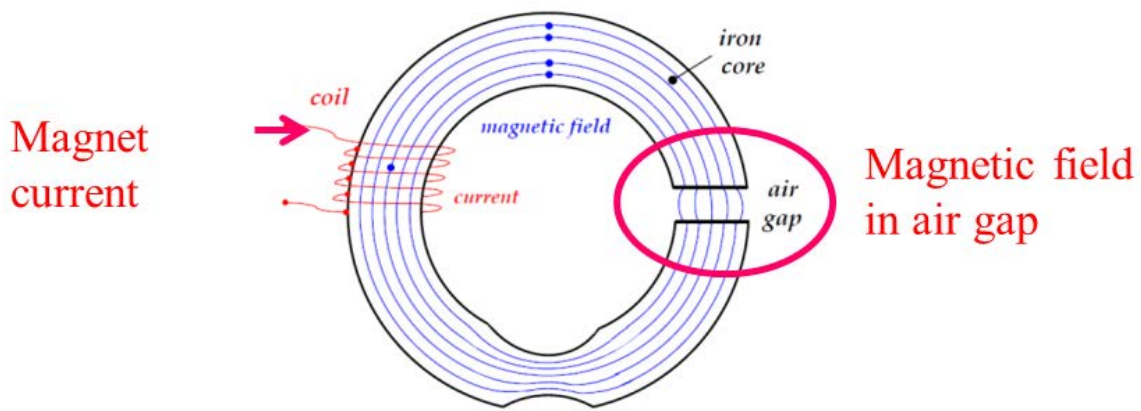

**Fig. 6:** Magnetic field

The main specification for the magnet power converters is high precision control of the current delivered to the magnets. Typically, the good field region of a magnet is defined within $\pm 10^{-4}$ $\Delta B/B$ and the typical performance of the power converter is to control the current to the order of $10^{-5}$ of the maximum output current.

The main difficulty is the fact that in a magnet the relation between the current and magnetic field is not linear due to magnetic hysteresis of the core and the eddy currents, see Fig. 7.

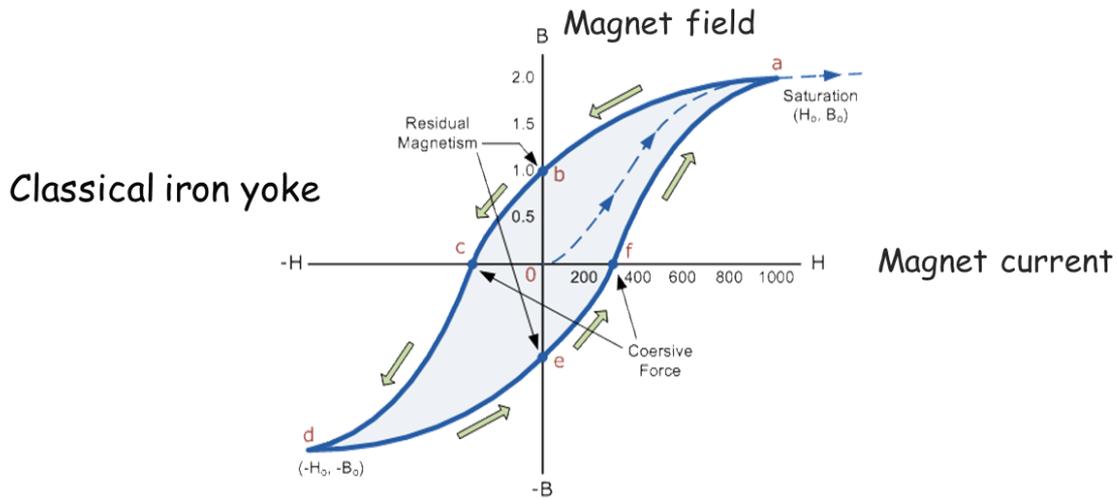

**Fig. 7:** Magnet hysteresis

This problem is one of the major difficulties when operating a synchrotron that is to work with different beam energies.

## 3.1 Magnet parameters

A good model of the load is needed to control the current delivered by the power converters. The transfer function gives the main parameters of the circuit which are inductance and resistance. To improve control, it is mandatory to include the saturation curve of the magnet to adapt the current loop parameters depending of the current level, see Fig. 8(b).

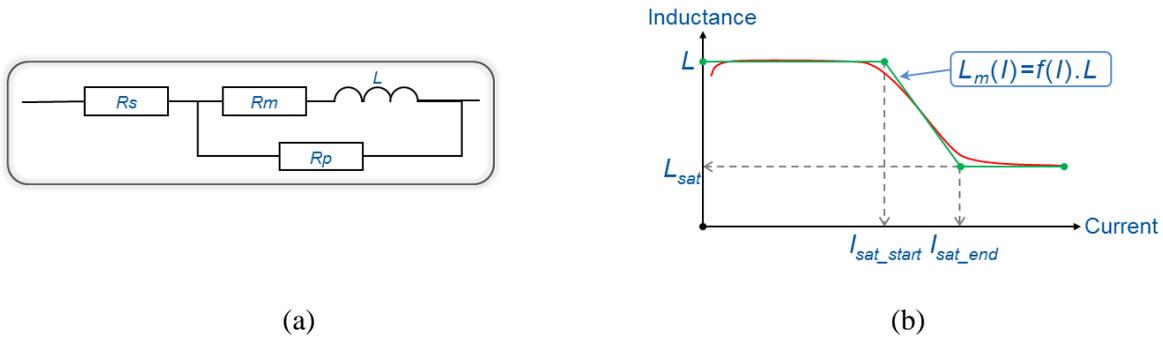

(a)                                                        (b)

**Fig. 8:** Magnet model: (a) equivalent circuit of the load (b) inductance value versus current

### 3.2 Circuit layout

The magnets can be powered individually or in series. The main advantage to powering them individually is the increase in flexibility of the beam optics. The main drawback is the uncertainty of the magnetic field between the magnets due to the magnet current history (hysteresis effect). The global cost is also higher, with more DC cables and more power converters. In some cases it can become mandatory to split a circuit because of the size of the load becoming too large. For example, this is the case for the LHC superconducting dipoles and quadrupoles, where the total energy of the circuit is very huge (8 GJ). The machine was therefore divided into eight sectors, see Fig. 9. Even with this split, the inductance of one octant is 15 H with a stored energy of more than 1 GJ. This configuration needs high-precision control of the magnet current as well as excellent tracking of the magnet current between the eight sectors [2].

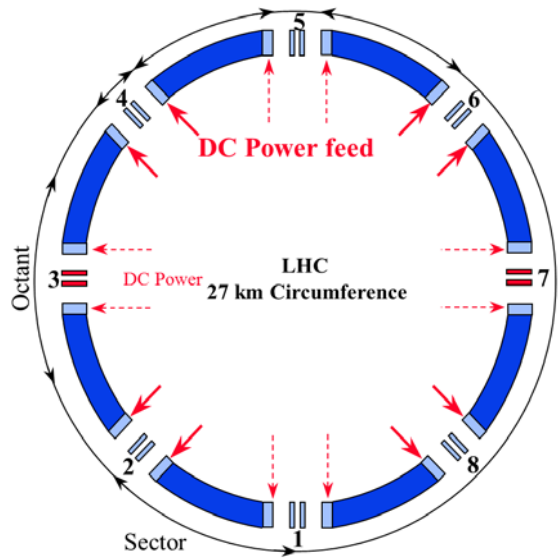

**Fig. 9:** Split of the LHC dipole and quadrupole circuits

The classical solution for dipole, quadrupole, and sextupole magnets is to connect all of them in series. The main advantage is that this suppresses the uncertainty of the magnetic field between magnets and assures easy control of these circuits. In some cases, trim power converters are needed to adapt the magnetic field locally. This solution is globally the cheapest one, but it requires a large power converter to power such circuits. As an example, the SPS machine requires a power system of 150 MW to power all of the dipole in series. A special distribution of power sources was chosen to reduce the common mode voltage applied to the magnets, see Fig. 10.

In some cases, the creation of nested circuits reduces the cost of the powering system. In a series of magnets that are powered by a main power converter, some magnets have additional connections to a trim power converter to adjust their current, see Fig. 11. This creates some difficulties for the power converter control as the different power converters are coupled by their load. This type of scheme isn't recommended due to the induced complexity. However, in some case the savings are such that it is difficult to avoid this solution [3].

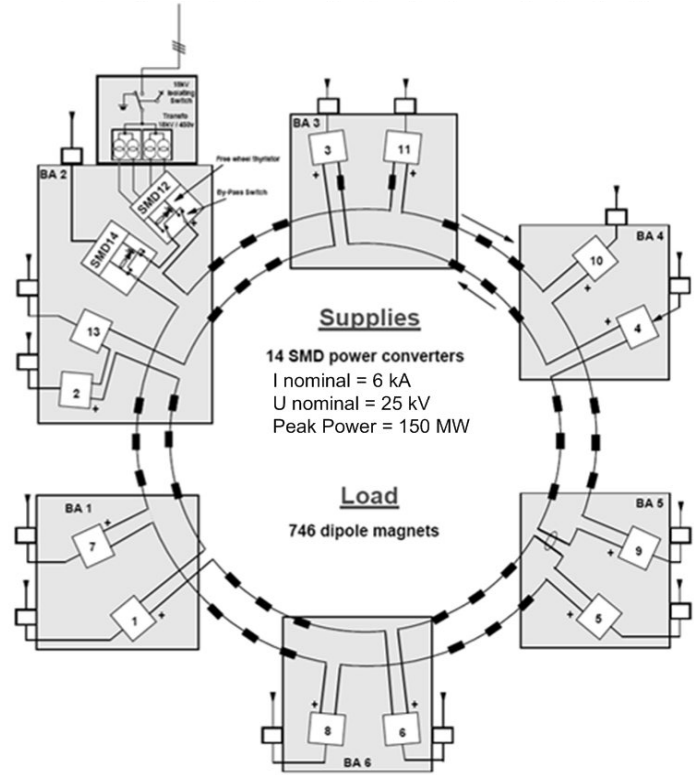

**Fig. 10:** SPS dipole power system

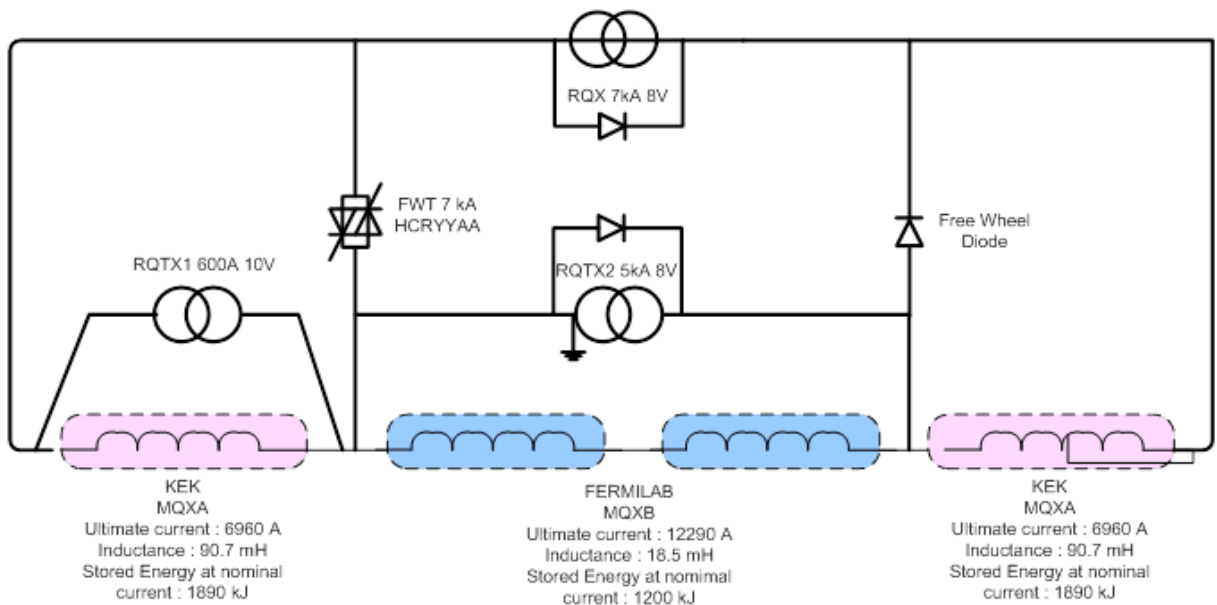

**Fig. 11:** LHC inner triplet powering scheme

## 3.3 Powering optimization

Powering optimization plays with the magnet parameters, the power converter parameters, and the circuit layout. For the same integral magnetic field, the magnet can be laid out in different ways. The magnet parameters are the number of turns per coil, the maximum current, the current density in the conductor, and the length of the magnet [4].

The number of turns per coil does not influence the total power losses in the magnet. By increasing the number of turns, the current required decreases linearly. This reduces the losses in the DC cables and power converters. The drawback is the higher voltage applied to the magnets and the increased size of the magnet due to insulation.

The magnet's current density determines the magnet's losses. The choice is made based on economic criteria. High current densities result in a small conductor cross-section, but with water cooling. This reduces the size of the magnet and thus the capital costs, but it requires larger power converters and gives increased running costs due to the magnet's losses. By lowering the current density, the electricity bills can be strongly reduced but at the price of having to invest in larger magnets, as shown in Fig. 12. A global optimization has to be carried out to find the best economical solution for the magnets as well as for the power converters.

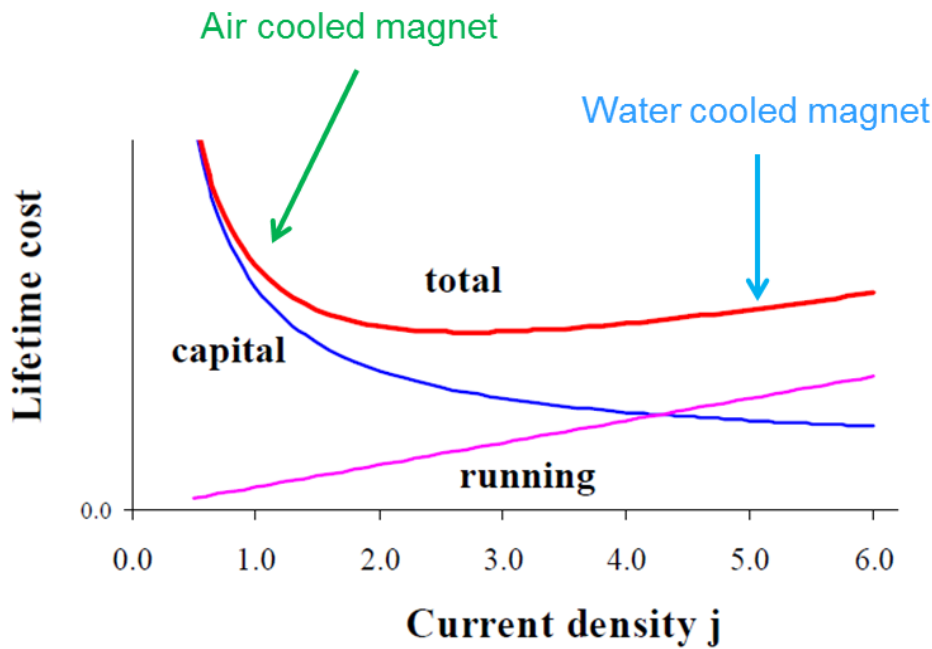

**Fig. 12:** The variation in magnet capital cost, running cost, and total lifetime cost versus conductor current density.

Another way to reduce power consumption is to reduce the time during which the magnets are powered. This is, for example, the case for linacs and transfer lines where the beam isn't always present, see Fig. 13. The idea is to power the magnets only when the beam is present, and this has a strong impact on electricity consumption. For example, in Linac4, the beam is pulsing at 2 Hz while the beam is present for 2 ms. The duty is then 0.4%. By pulsing the magnets, when compared to DC magnets, the savings are enormous (99%). This technique requires special designs for those power converters where energy storage can be implemented.

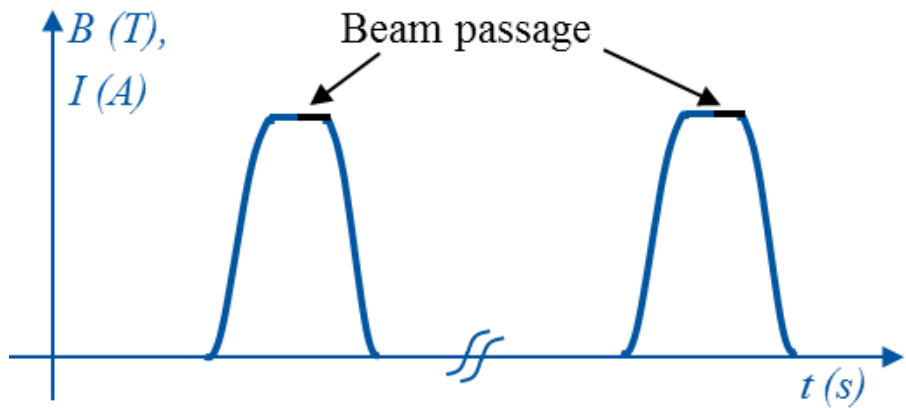

**Fig. 13:** Beam passage versus time

### 3.4 Magnet grounding

For safety reasons, magnets are isolated from the electrical grid. The power converter needs an isolation transformer in its topology. Magnets are connected to the ground at one point; they can't be left floating with their parasitic capacitances. One polarity can be connected directly to the ground through a resistor to limit the earth fault current, or via a divider for better voltage sharing, see Fig. 14. The ground current is monitored and stops the power converter in the case of an earth fault in the DC circuit.

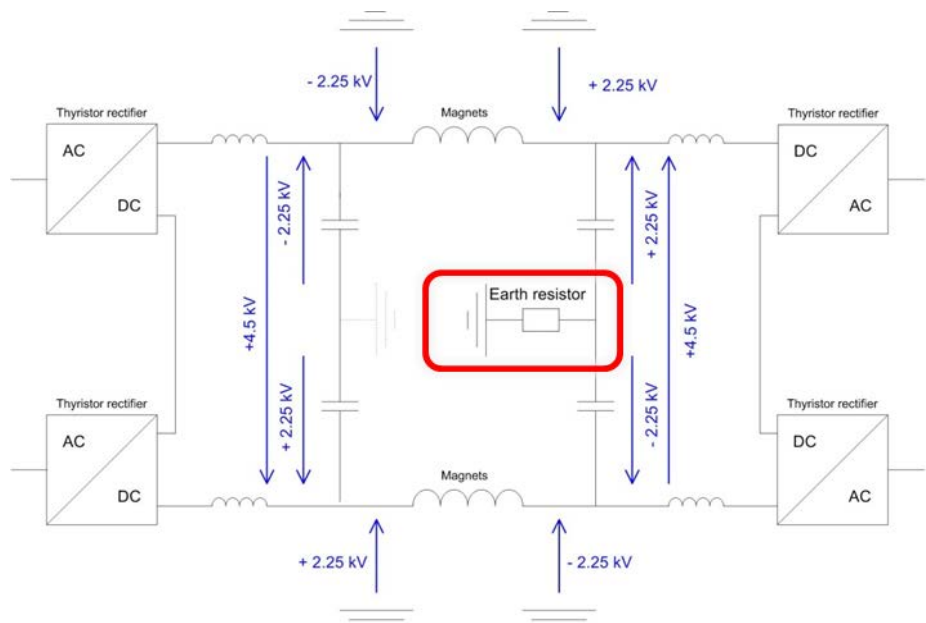

**Fig. 14:** An example of earthing the system

For a large number of magnets in series, the arrangement of multiple power sources can be made so that the common mode voltage of the magnets (any polarity of the magnet to ground) is reduced, when compared to the total applied voltage in differential mode. As shown in Fig. 14, the total applied voltage is 9 kV while the maximum common voltage of the magnets is limited to 2.25 kV.

## 3.5 Magnet protection

The magnets will have an interlock system that requests the power converter to stop in the event of any faults. For warm magnets, it is quite simple (water flow, thermostat, red button, etc.) For superconducting magnets, it is quite complex (quench protection, cryogenics, etc.)

In the event of a fault, the magnet current will be stopped but, because the magnets are inductive loads, the circuit can't be opened. The power converters assure a freewheeling path to the current, and the decay time is determined by the time constant of the circuit ($L$, $R$). In some cases, an additional dump resistor is needed to accelerate the discharge time, see Fig. 15.

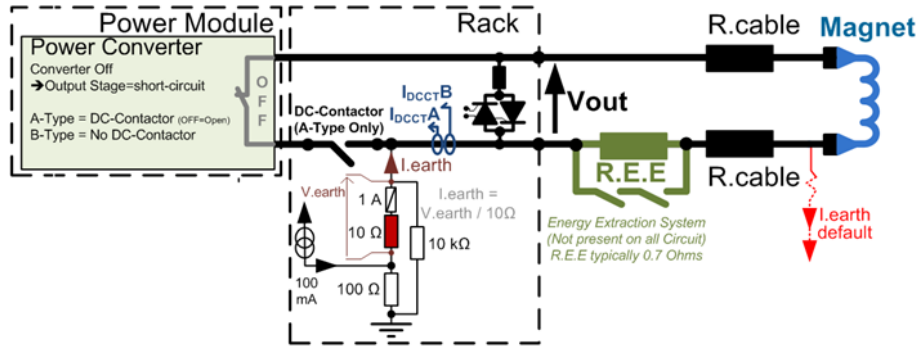

**Fig. 15:** Output circuit for magnet protection

## 3.6 Operation cycle

Magnet operation has a strong impact on the power converter topology and on power converter ratings. The magnet current cycle has to be defined from the beginning of a project, see Fig. 16.

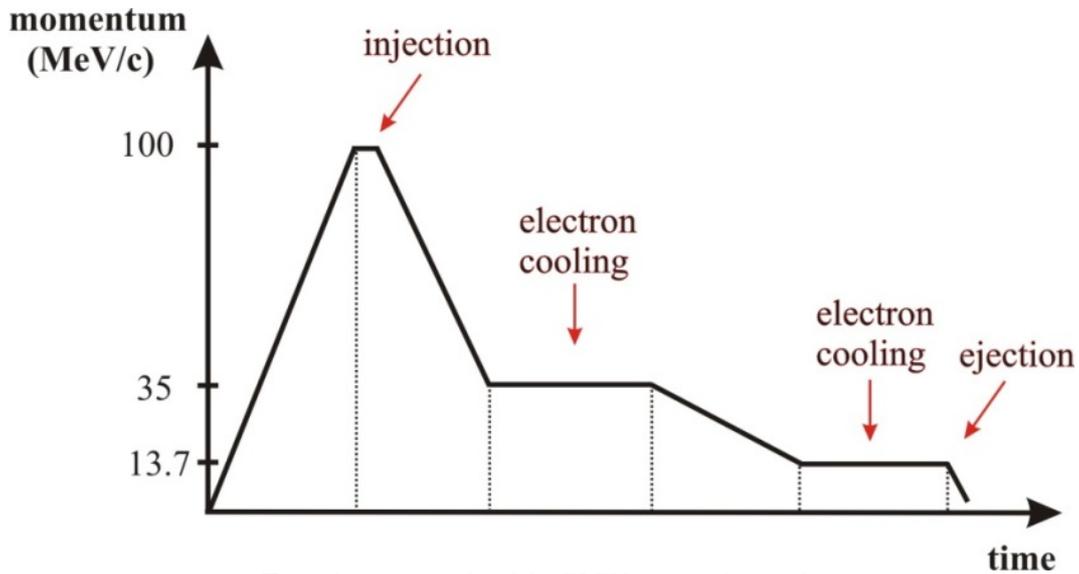

**Fig. 16:** An example of the ELENA operating cycle

The main parameters to be defined are the minimum and maximum operating current and the ramp rates. This will define the voltage needed and the topology of the converters. If the applied current and voltage are always positive, then a one-quadrant converter can be selected. If the current is always positive but the voltage is bipolar, then the power converter must be of the two-quadrant type. If the voltage and the current are bipolar, then, a four-quadrant power converter will be required, see Fig. 17.

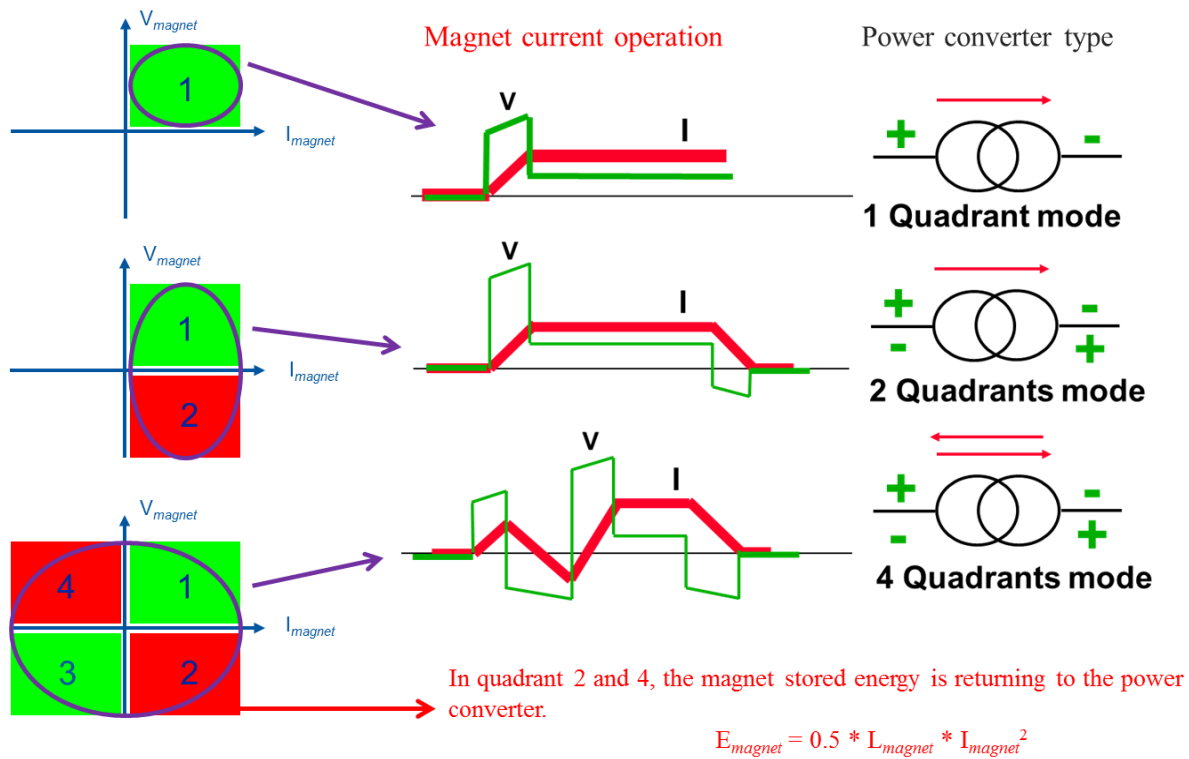

**Fig. 17:** The power converter type depends upon the operating cycle

## 4 Power converters

The power converter types will be introduced in the following papers. Control of power converter current is a challenge, and the main principles are described below.

### 4.1 Power converter control

Power converter performance has to be defined at the beginning of the project, based on the accelerator's requirements. The power converters needed for particle accelerators are always of high-class precision. The term 'precision' is only a generic term covering accuracy, reproducibility, and stability. For each power converter, the requirements depend on the magnet type and function. The most demanding are the dipole and quadrupole magnets, typically in the order of $1\text{–}10 \times 10^{-5}$, while the performance for corrector magnets are less demanding, typically in the order of $1\text{–}10 \times 10^{-4}$.

Power converters are always of the current control type, and the control principle is shown in Fig. 18. The power converter receives a current reference from the control room that needs to be executed with precise timing. All of the machine's magnets need to be synchronized, and a central timing system is distributed to all power converters and other accelerator devices. The precise execution of the current reference is one of the most challenging aspects of the control system.

The performance of the current control can be monitored through the tracking error, which is the ability of the power converter to follow the reference function. The static part of the tracking error is linked to static performance (accuracy and reproducibility). The dynamic part comes from timing error and regulation lagging error. All of these requirements lead to a definition for the power converter controller.

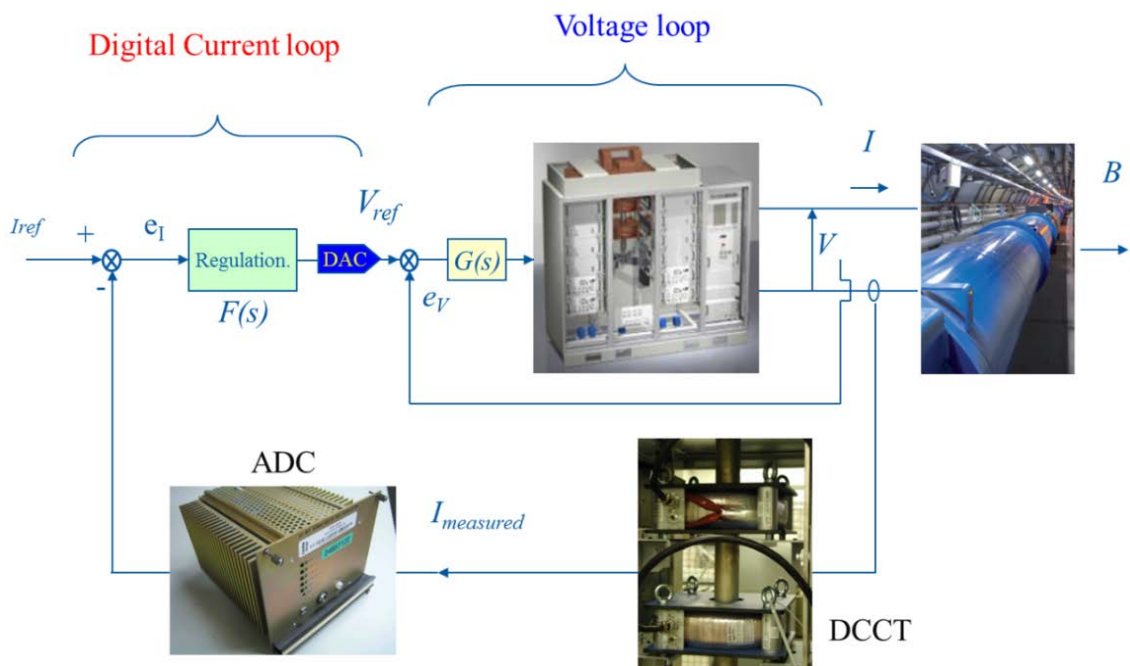

**Fig. 18:** Principle of power converter control

## 4.2 High precision

To get a high-precision power converter, a device capable of precisely measuring the output current is needed. The most popular current transducer is the DCCT (Direct-Current Current Transducer) for its high performance in many different fields, see Fig. 19. It is also classical to install two devices to be able to monitor any deviations of the currents between them [5].

|  | DCCTs | Hall effect | CTs | Rogowsky | Shunts |
|---|---|---|---|---|---|
| Principle | Zero flux detection | Hall effect | Faraday's law | Faraday's law | Ohm's law |
| Output | Voltage or current | Voltage or current | Voltage | Voltage | Voltage |
| Accuracy | Best devices can reach a few ppm stability and repeatability | Best devices can reach 0.1% | Typically not better than 1% | Typically %, better possible with digital integrators | Can reach a few ppm for low currents, <% for high currents |
| Ranges | 50A to 20kA | hundreds mA to tens of kA | 50A to 20kA | high currents possible, up to 100kA | From <mA up to to several kA |
| Bandwidth | DC ..kHz for the higher currents, DC..100kHz for lower currents | DC up to couple hundred kHz | Typically 50Hz up to a few hudreds of kHz | Few Hz possible, up to the MHz | Up to some hundreds of kHz with coaxial assemblies |
| Isolation | Yes | Yes | Yes | Yes | No |
| Error sources | Magnetic (remanence, external fields, centering) Burden resistor (thermal settling, stability, linearity, tempco) Output amplifier (stability, noise, CMR, tempco) | Magnetic Burden resistor Output amplifier Hall sensor stability (tempco, piezoelectric effect) | Magnetic (remanence, external fields, centering, magnetizing current) Burden resistor | Magnetic Integrator (offset stability, linearity, tempco) | Power coefficient, tempco, ageing, thermal voltages |

**Fig. 19**: Different types of current transducers

The performance of the power converter also strongly depends upon the current regulation control algorithm. The most powerful algorithm is the RST controller, which allows management of the tracking error as well as regulation, see Fig. 20 [6].

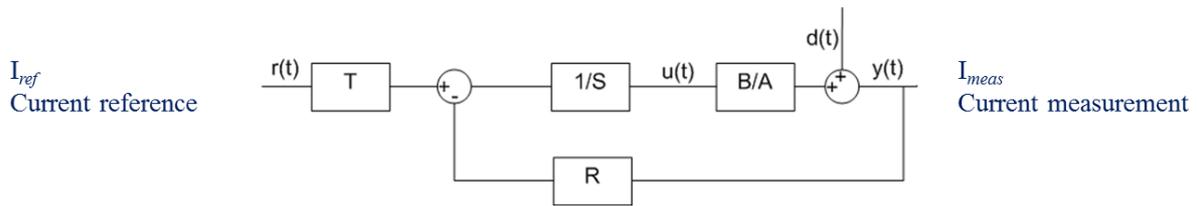

**Fig. 20:** RST controller

### 4.3 Power converter ripple

In principle, the voltage delivered by a power converter has a ripple that is due to semiconductor switching. The voltage ripple is generated by the power converter, and it is converted into a magnetic field ripple through the impedance of the magnet. The maximum magnetic field ripple has to be determined by the beam quality requirements, and then from the impedance of the magnet the maximum voltage ripple can be specified, see Fig. 21. The impedance of the vacuum chamber, between the current ripple and the magnetic field ripple, can also be taken into account. Typically, the vacuum chamber has a filtering effect above 100 Hz. The voltage ripple has to be specified for all frequencies, see Fig. 22.

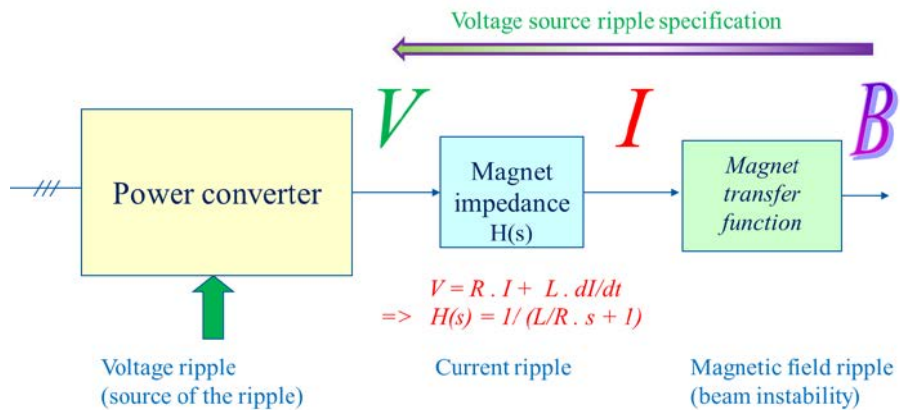

**Fig. 21:** Ripple chain

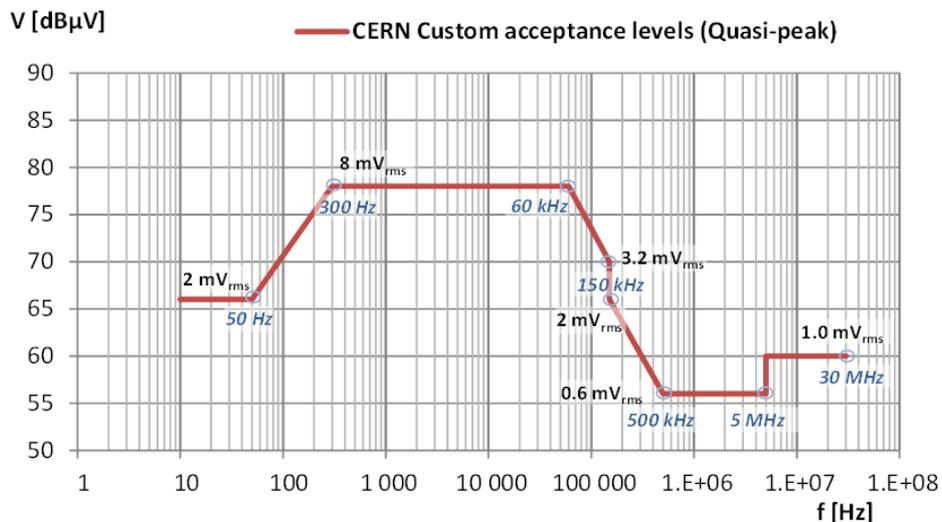

**Fig. 22:** Example of voltage ripple specification

## 4.4 Functional specification

Before starting the design of a new power converter, a functional specification needs to be approved between the accelerator physicists and the magnet designers. This functional specification includes:

- a short description of the machine;
- a description of the loads: magnet layout, magnet parameters, optimization with integral cost, and energy saving;
- a description of the operation duty cycle: machine cycles, minimum and maximum beam energy, ramp rates, hysteresis management, etc.;
- power converter requirements: power converter rating, current precision, current tracking, control system, energy management, lock-out and safety procedure, infrastructure (layout, electricity, cooling, handling, etc.);
- purchasing and development strategy;
- planning;
- budget;
- resources.

## 5 Introduction to the main challenges

The design of power converters covers a large range of disciplines. It needs more than one specialist to build it, it is a team work. The main challenges are listed below:

- power:
    - power converter topologies;
    - semiconductors, switching frequency, thermal design, fatigue while cycling…
    - filtering and electromagnetic compatibility (EMC);
    - connection to AC grid and robustness to grid perturbations;
    - energy management and energy saving;
    - protection and safety of the system;
- control:
    - accuracy class;
    - digitalization;
    - control loops;
    - timing and synchronization;
    - control interfaces;
    - interlocks with external systems.